\documentclass[twocolumn,A4]{article}
\usepackage[dvips]{graphics}
\usepackage{setspace}
\usepackage{color}
\definecolor{gold}{rgb}{0.85,0.66,0}
\definecolor{dblue}{rgb}{0,0,0.8}
\begin{document}
\onecolumn
\begin{center}
{\bf{\Large {\textcolor{gold}{Tuning of electron transport through a 
moebius strip: shot noise}}}}\\
~\\
{\textcolor{dblue}{Santanu K. Maiti}}$^{1,2,*}$ \\
~\\
{\em $^1$Theoretical Condensed Matter Physics Division,
Saha Institute of Nuclear Physics, \\
1/AF, Bidhannagar, Kolkata-700 064, India \\
$^2$Department of Physics, Narasinha Dutt College,
129, Belilious Road, Howrah-711 101, India} \\
~\\
{\bf Abstract}
\end{center}
We explore electron transport through a moebius strip attached to two 
metallic electrodes by the use of Green's function technique. A parametric 
approach is used based on the tight-binding model to characterize the 
electron transport through such a bridge system and it is observed that 
the transport properties are significantly affected by (a) the transverse 
hopping strength between the two channels and (b) the strip-to-electrode 
coupling strength. In this context we also describe the noise power of 
the current fluctuations that provides a key information about the electron 
correlation which is obtained by calculating the Fano factor ($F$). The 
knowledge of this current fluctuations gives important ideas for 
fabrication of efficient electronic devices.
\vskip 1cm
\begin{flushleft}
\end{flushleft}
\vskip 4.5in
\noindent
{\bf ~$^*$Corresponding Author}: Santanu K. Maiti

Electronic mail: santanu.maiti@saha.ac.in
\newpage
\twocolumn

\section{Introduction}

The advancements in nanoscience and technologies prompting a growing number 
of researchers across multiple disciplines to attempt to devise innovative 
ways for decreasing the size and increasing the performance of microelectronic 
circuits. One possible route is based on the idea of using molecules and 
molecular structures as functional devices. Following experimental 
developments, theory can play a major role in understanding the new 
mechanisms of conductance, but, the goal of developing a reliable 
molecular-electronics technology is still over the horizon and many key 
problems, such as device stability, reproducibility and the control of 
single-molecule transport need to be solved. Quantum transport properties 
through molecules was first studied theoretically during 
$1970$'s~\cite{aviram}. Later several experiments~\cite{metz,fish,reed1,
reed2,tali} have been carried out on electron transport through molecules 
placed between two metallic electrodes with few nanometer separation.
It is very essential to control electron conduction through such quantum 
devices and the present understanding about it is quite limited. For example, 
it is not so clear how the molecular transport is affected by the structure 
of the molecule itself or by the nature of its coupling to the 
electrodes~\cite{xue1}. The electron conduction through such two-terminal 
devices can be controlled by some bias or gate voltage across the device. 
The current passing across the junction then becomes a strongly non-linear 
function of the applied voltage, and its understanding is a highly 
challenging problem. The knowledge of current fluctuations (quantum origin) 
provides a key idea for fabrication of efficient molecular devices.
Blanter {\em et al.}~\cite{butt} have described elaborately how the lowest 
possible noise power of the current fluctuations can be determined in a 
two-terminal conductor. The steady state current fluctuations, the so-called 
shot noise, is a consequence of the quantization of charge and it can be used
to obtain information on a system which is not available through conductance
measurements. The noise power of the current fluctuations provides an 
additional important information about the electron correlation by calculating 
the Fano factor ($F$) which directly informs us whether the magnitude of the 
shot noise achieves the Poisson limit ($F=1$) or the sub-Poisson ($F<1$) 
limit.

In a recent experiment Tanda {\em et al.}~\cite{tanda} have fabricated a 
microscopic NbSe$_3$ moebius strip and it raises some interesting questions 
regarding the topological effect on quantum transport. In some theoretical 
papers~\cite{avishai,weiden} the topological effect on quantum transport for 
isolated moebius strips has been described and it has been observed that 
the transverse hopping strength has significant role on such transport.
For non-zero transverse hopping strength, a moebius strip becomes a regular
two-channel ring. On the other hand, when electrons are unable to move
along the transverse direction a moebius strip reduces to a single-channel
ring with doubling its length than the previous one. This is due to the
peculiar topology of a moebius strip. In the present article we are 
interested about the electron transport properties through such a moebius 
strip which is attached to two metallic electrodes. Here we address a 
simple analytical formulation of the transport problem through the moebius 
strip using the tight-binding Hamiltonian. There exist some {\em ab initio} 
methods for the calculation of conductance~\cite{yal,ven,xue,tay,der,dam,
bran,rocha} as well as model calculations~\cite{muj1,muj2,sam,hjo,baer1,
baer2,baer3,walc1,walc2}. The model calculations are motivated by the fact 
that the {\em ab initio} methods are computationally too expensive and 
also provide a better insight to the problem and here we concentrate only 
on the qualitative results rather than the quantitative ones.

The plan of the paper is as follows. Section $2$ describes the methodology 
for the calculation of electron transport through a finite size conductor 
sandwiched between two metallic electrodes. In Section $3$, we study the 
conductance ($g$) behavior as a function of energy ($E$), the current ($I$)
and the noise power ($S$) of its fluctuations as a function of applied 
bias voltage ($V$) for some typical moebius strips. Finally, we summarize 
our results in Section $4$.

\section{Theoretical formulation}

\begin{figure}[ht]
{\centering \resizebox*{7.5cm}{1.5cm}{\includegraphics{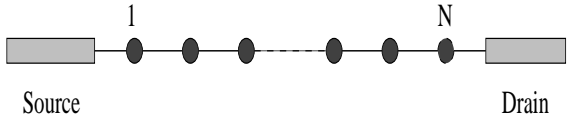}}\par}
\caption{Schematic representation of a bridge system. A one-dimensional 
conductor with $N$ atomic sites attached to two metallic electrodes, 
namely, source and drain.}
\label{dot}
\end{figure}

Here we formulate the technique for the calculation of transmission 
probability ($T$), conductance ($g$), current ($I$) and noise power 
of current fluctuations ($S$) for a finite size conductor attached 
with two metallic electrodes (schematically represented in Fig.~\ref{dot}) 
by the use of Green's function technique.

At low temperatures and bias voltage the conductance $g$ of the conductor 
is given by the Landauer conductance formula~\cite{datta},
\begin{equation}
g=\frac{2e^2}{h} T
\label{equ1}
\end{equation}
where the transmission probability $T$ becomes~\cite{datta},
\begin{equation}
T={\mbox{Tr}}\left[\Gamma_S G_C^r \Gamma_D G_C^a\right]
\label{equ2}
\end{equation}
where $G_C^r$ ($G_C^a$) is the retarded (advanced) Green's function of the
conductor and $\Gamma_S$ ($\Gamma_D$) describes its coupling to the
source (drain). The Green's function of the conductor is expressed as,
\begin{equation}
G_C=\left(E-H_C-\Sigma_S-\Sigma_D\right)^{-1}
\label{equ3}
\end{equation} 
where $E$ is the energy of the injecting electron and $H_C$ is the Hamiltonian 
of the conductor which can be expressed in the tight-binding model within the 
non-interacting picture as,
\begin{equation}
H_C=\sum_i \epsilon_i c_i^{\dagger} c_i + \sum_{<ij>} t \left(c_i^{\dagger} 
c_j + c_j^{\dagger} c_i\right)
\label{equ4}
\end{equation}
where $\epsilon_i$'s are the site energies and $t$ is the nearest-neighbor 
hopping strength. In Eq.(\ref{equ3}), $\Sigma_S$ and $\Sigma_D$ correspond to 
the self-energies due to the coupling of the conductor to the source and
drain, respectively. All the information regarding the electrode-to-conductor 
coupling are included into these two self-energies and are described through 
the use of Newns-Anderson chemisorption theory~\cite{muj1,muj2}. 

The current passing through the conductor can be considered as a single 
electron scattering process between the two reservoirs of charge carriers. 
The current-voltage relation can be obtained from the expression~\cite{datta},
\begin{equation}
I(V)=\frac{e}{\pi \hbar}\int \limits_{-\infty}^{\infty} \left(f_S-f_D\right) 
T(E) dE
\label{equ5}
\end{equation}
where the Fermi distribution function $f_{S(D)}=f\left(E-\mu_{S(D)}\right)$ 
with the electrochemical potentials $\mu_{S(D)}=E_F\pm eV/2$. Here we assume,
for the sake of simplicity, that the entire voltage is dropped across the 
conductor-to-electrode interfaces and this assumption does not greatly affect
the qualitative features of the current-voltage characteristics.

The noise power of the current fluctuations is calculated from the following
expression~\cite{butt},
\begin{eqnarray}
S & = & \frac{2e^2}{\pi \hbar}\int \limits_{-\infty}^{\infty}\left[T(E)
\left\{f_S\left(1-f_S\right) + f_D\left(1-f_D\right) \right\} \right. 
\nonumber \\ 
 & & + T(E) \left. \left\{1-T(E)\right\}\left(f_S-f_D\right)^2 \right] dE
\label{equ6}
\end{eqnarray}
where the first two terms of this equation correspond to the equilibrium 
noise contribution and the last term gives the non-equilibrium or shot noise
contribution to the power spectrum. Calculating the noise power we can 
compute the Fano factor $F$, which is essential to predict whether the shot 
noise lies in the Poisson or the sub-Poisson limit, through the 
relation~\cite{butt},
\begin{equation}
F=\frac{S}{2 e I}
\label{equ7}
\end{equation}
For $F=1$, the shot noise achieves the Poisson limit where no electron 
correlation exists between the charge carriers. On the other hand for $F<1$, 
the shot noise reaches the sub-Poisson limit and it provides the information 
about the electron correlation among the charge carriers.

In this article we focus our results for a moebius strip at very low 
temperature ($5$ K), but the qualitative features of all the results are 
invariant up to some finite temperature ($\sim 300$ K). For simplicity we 
take the unit $c=e=h=1$ in our all calculations.

\section{Results and their interpretation}

\begin{figure}[ht]
{\centering \resizebox*{7cm}{3.5cm}{\includegraphics{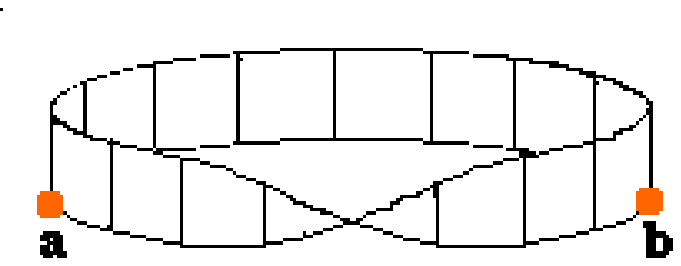}}\par}
\caption{Schematic view of a moebius strip. The two electrodes are attached 
to the strip at the points $a$ and $b$ (colored spots), respectively.} 
\label{moebius}
\end{figure}

The schematic geometry of a moebius strip is shown in Fig.~\ref{moebius}.
The two metallic electrodes are attached to the strip at the points $a$
and $b$ (colored spots in Fig.~\ref{moebius}), respectively. In experimental 
set up, these electrodes are constructed from the gold (Au) atoms and the 
bridging system is attached to the electrodes by thiol (S-H) groups in 
the chemisorption technique where the hydrogen (H) atoms remove and the 
sulfur (S) atoms reside. The tight-binding Hamiltonian that describes 
the moebius strip with $M$ rungs is written in this form, 
\begin{eqnarray}
H_C &=& \sum_{i=1}^{2M}\epsilon_i c_i^{\dagger}c_i + t\sum_{i=1}^{2M}\left(
c_i^{\dagger}c_{i+1} + c_{i+1}^{\dagger}c_i \right) \nonumber \\
& & +~ t_{\bot}\sum_{i=1}^{2M} c_i^{\dagger}c_{i+M}
\label{equ8}
\end{eqnarray}
where $t_{\bot}$ corresponds to the transverse hopping strength between the
two channels, $t$ represents the nearest-neighbor hopping strength along the 
\begin{figure}[ht]
{\centering \resizebox*{8cm}{10cm}{\includegraphics{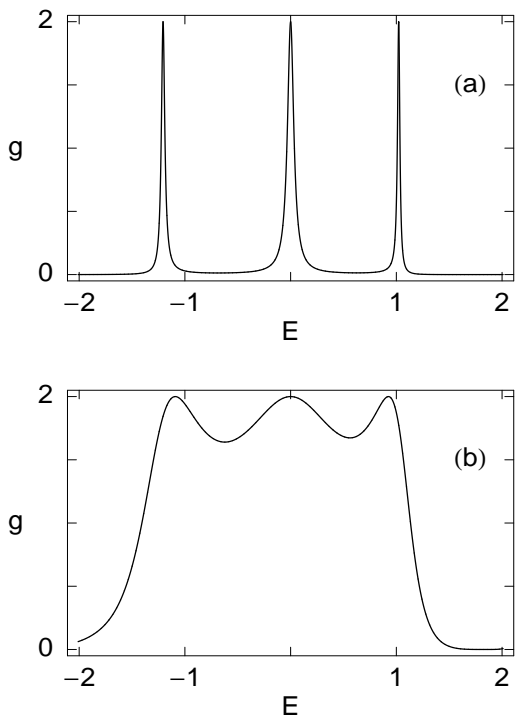}}\par}
\caption{Conductance ($g$) as a function of the injecting electron energy 
($E$) for the moebius strip with $t_{\bot} \ne 0$. For this case the strip
can be treated as a regular two-channel ring. Here we take the total number
of rungs $M=6$ (even). (a) and (b) correspond to the results for the weak 
and the strong coupling limits, respectively.} 
\label{finitecond}
\end{figure}
longitudinal direction and all the other symbols carry the same meaning as in 
Eq.~\ref{equ4}. This strip has a special kind of geometry and the electron
transport is strongly affected by the transverse hopping strength ($t_{\bot}$).
Not only that, the transport characteristics are also significantly 
influenced by the strip-to-electrode coupling strength ($\tau_{\{S,D\}}$), 
and, here we focus our results in these two aspects. 

We shall describe the electron transport characteristics through the moebius 
strip in the two different regimes depending on both (a) the transverse
hopping strength between the two channels and (b) the coupling strength 
of the strip to the electrodes. For the transverse hopping strength, we 
take the two distinct regimes in the following way. One is the case where 
the electrons are able to hop between the two channels i.e., $t_{\bot} 
\ne 0$. Here we take the value of $t_{\bot}$ as identical with the value 
of $t$, for simplicity. In the other case the electrons can not move along 
the transverse direction i.e., 
\begin{figure}[ht]
{\centering \resizebox*{8cm}{10cm}{\includegraphics{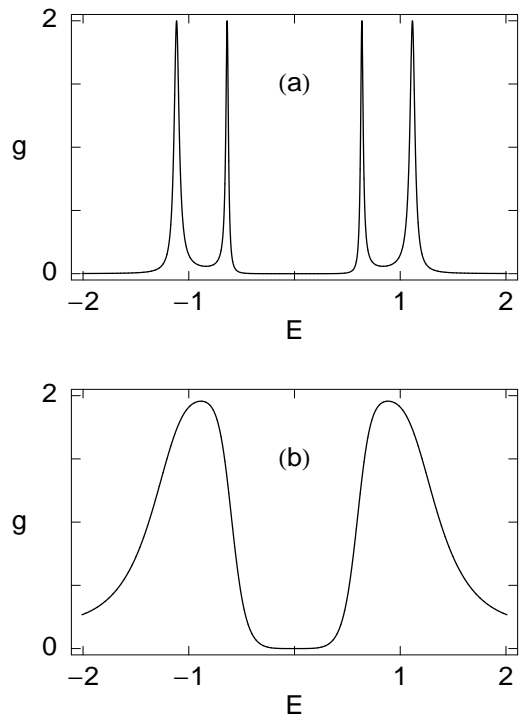}}\par}
\caption{Conductance ($g$) as a function of the injecting electron energy 
($E$) for the moebius strip with $t_{\bot}=0$. For this case the moebius
strip reduces to a single chain of length $2M$. Here we take $M=6$ (even)
so that the length of the chain is $12$ in unit of the lattice spacing.
(a) and (b) correspond to the results for the weak and the strong coupling 
limits, respectively.} 
\label{zerocond}
\end{figure}
$t_{\bot}=0$. The values of the parameters for these two separate cases are 
taken as: $t=t_{\bot}=2.5$ and $t=2.5$, $t_{\bot}=0$, respectively. On the
other hand depending on the strip-to-electrode coupling strength we classify 
the two distinct regimes as follows. One is the so-called weak-coupling 
regime $\tau_{S(D)} << t$ and the other one is the strong-coupling regime 
$\tau_{S(D)} \sim t$, where $\tau_S$ and $\tau_D$ are the hopping strengths 
of the strip to the source and drain, respectively. In our calculations, 
the parameters in these two regimes are chosen as $\tau_S=\tau_D=0.5$, 
$t=2.5$ (weak-coupling) and $\tau_S=\tau_D=2$, $t=2.5$ (strong-coupling). 
Here we set the site energy $\epsilon_i=0$ for all the sites of the strip.

In Fig.~\ref{finitecond}, we plot the conductance $g$ as a function of the
injecting electron energy $E$ for the moebius strip when the electrons are
\begin{figure}[ht]
{\centering \resizebox*{8cm}{10cm}{\includegraphics{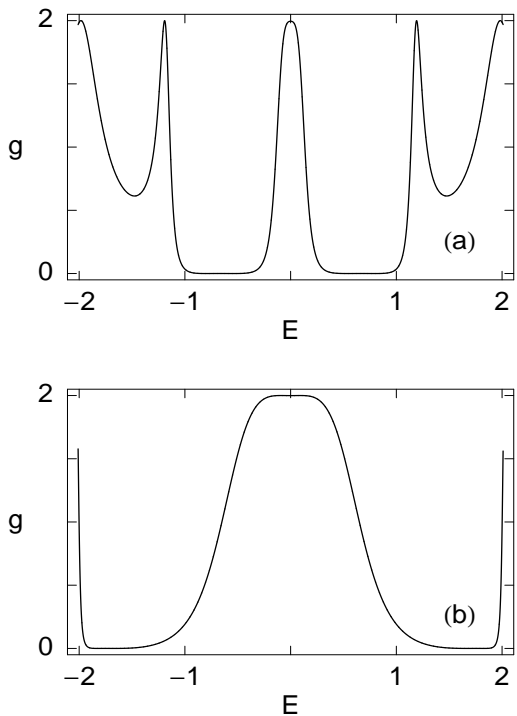}}\par}
\caption{Conductance ($g$) as a function of the injecting electron energy 
($E$) for the moebius strip with $t_{\bot} \ne 0$. For this case the strip
can be treated as a regular two-channel ring. Here we take the total number
of rungs $M=7$ (odd). (a) and (b) correspond to the results for the weak 
and the strong coupling limits, respectively.} 
\label{finitecondodd}
\end{figure}
allowed to hop along the transverse direction i.e., $t_{\bot} \ne 0$. For 
such a case, the moebius strip eventually becomes a small cylinder and no
matter it is whether the system is twisted or not. Therefore, for this case
the moebius strip can be treated as a regular two-channel ring. 
Figure~\ref{finitecond}(a) and (b) correspond to the results of the strip 
for the weak and the strong coupling limits, respectively. In the limit of 
weak coupling, the conductance shows sharp resonant peaks 
(Fig.~\ref{finitecond}(a)) for some specific energy values, while, for all 
other energies it drops to zero. At these resonances, the conductance gets 
the value $2$ and accordingly, the transmission probability $T$ goes to 
unity (since from the Landauer conductance formula we get the relation 
$g=2T$ (Eq.~\ref{equ1}) as $e=h=1$ in our present calculations).
These resonant peaks are associated with the energy eigenvalues of the 
\begin{figure}[ht]
{\centering \resizebox*{8cm}{10cm}{\includegraphics{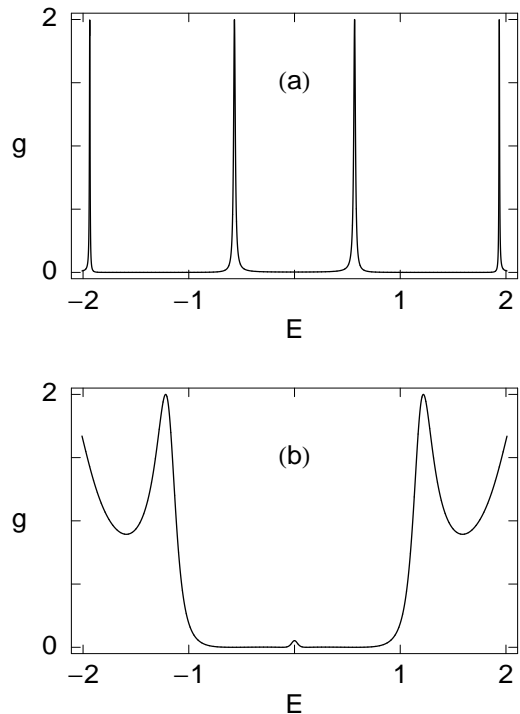}}\par}
\caption{Conductance ($g$) as a function of the injecting electron energy 
($E$) for the moebius strip with $t_{\bot}=0$. For this case the moebius
strip reduces to a single chain of length $2M$. Here we take $M=7$ (odd)
so that the length of the chain is $14$ in unit of the lattice spacing.
(a) and (b) correspond to the results for the weak and the strong coupling 
limits, respectively.} 
\label{zerocondodd}
\end{figure}
moebius strip. Therefore the conductance spectrum manifests itself the 
energy eigenvalues of the strip. Now for the strong coupling limit, the
widths of these resonant peaks get substantial broadening as observed from
Fig.~\ref{finitecond}(b). This feature appears due to the fact that the
energy levels of the strip become broadened in the limit of strong coupling.
The contribution for this broadening of the energy levels comes from the
imaginary parts of the two self energies $\Sigma_S$ and 
$\Sigma_D$~\cite{datta}. From the results shown in Figs.~\ref{finitecond}(a)
and (b), we see that in the weak coupling limit the conduction of the 
electron through the bridge system takes place for very sharp energy ranges, 
while, in the limit of strong coupling the conduction takes place almost for 
all energy ranges. Therefore, by tuning the coupling strength the electron 
conduction through the strip can be controlled efficiently.

Figure~\ref{zerocond} shows the conductance variation as a function of the
energy $E$ for the moebius strip when the electrons are unable to hop along 
the transverse direction i.e., $t_{\bot}=0$. For this particular case 
($t_{\bot}=0$), an electron that moves along the longitudinal direction of
the loop encircles the loop twice before returning its starting point.
\begin{figure}[ht]
{\centering \resizebox*{8cm}{9.5cm}{\includegraphics{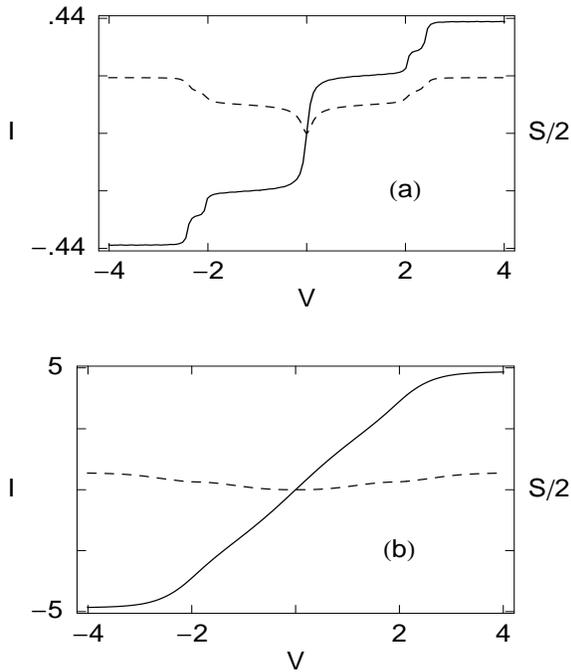}}\par}
\caption{Current $I$ (solid line) and the noise power of its fluctuations 
$S$ (dotted line) as a function of the voltage $V$ for the moebius strip 
with $t_{\bot} \ne 0$. For this case the moebius strip is equivalent to a
two-channel ring. Here we set the total number of rungs $M=6$ (even). (a) 
and (b) correspond to the results for the weak and the strong coupling 
limits, respectively.}
\label{finitecurr}
\end{figure}
Accordingly, it traverses twice path length i.e., the effective length 
becomes $2M$ and then the system is treated as a single chain. Here we 
set $M=6$, to compare the results with the two-channel ring (plotted in 
Fig.~\ref{finitecond}), so that the length of the chain becomes $12$ ($2M$)
in unit of the lattice spacing. The results for the weak and the strong 
coupling limits are given in Figs.~\ref{zerocond}(a) and (b), respectively. 
Similar to the results as plotted in Fig.~\ref{finitecond}, here we also 
get the sharp resonant peaks in the conductance spectrum for the weak 
coupling case, while, these resonances get broadened in the limit 
of strong coupling. The significant observation is that in this bridge 
system, both for the weak and strong coupling limits, electron conduction 
starts beyond some finite value of the energy $E$ (contrary to the results 
as given for the moebius strip with finite transverse hopping 
(Fig.~\ref{finitecond}), where a sharp resonant peak is observed across 
the energy $E=0$). Therefore, the results for the moebius strip with 
$t_{\bot}=0$ (Fig.~\ref{zerocond}) predict that the electron conducts 
through the system beyond some critical value of the applied bias voltage. 

In order to reveal the dependence of the conductance behavior on the total 
number of rungs $M$, now we describe the results for the moebius strip with
odd number of rungs. As illustrative example, in Fig.~\ref{finitecondodd}
we plot the conductance $g$ as a function of the injecting electron energy
$E$ for the moebius strip with non-zero transverse hopping strength i.e.,
$t_{\bot} \ne 0$. Here we set the total number of rungs $M=7$. The results
for the weak and the strong coupling cases are shown in 
Figs.~\ref{finitecondodd}(a) and (b), respectively. Similar to the results
observed in Fig.~\ref{finitecond}, here we also see that the widths of
the resonant peaks get broadened as the coupling strength goes from the
weak (Fig.~\ref{finitecondodd}(a)) to the strong (Fig.~\ref{finitecondodd}(b)) 
limit. Both for the weak and strong coupling cases, we get a sharp resonant 
peak across the energy $E=0$ (similar to the results as described in 
Fig.~\ref{finitecond}).

In Fig.~\ref{zerocondodd}, we plot the characteristic behavior of the 
conductance $g$ as a function of the energy $E$ for the moebius strip ($M=7$) 
when the electrons are unable ($t_{\bot}=0$) to hop along the transverse 
direction. Figure~\ref{zerocondodd}(a) and (b) correspond to the results for 
the weak and strong coupling cases, respectively. These features are almost 
quite similar to the results predicted for the moebius strip with even 
($M=6$) number of rungs considering $t_{\bot}=0$ (see Fig.~\ref{zerocond}).
From the results given in Fig.~\ref{zerocondodd}, we see that the electron
starts conduction across the strip beyond some finite value of the energy 
$E$, similar to the results as observed in Fig.~\ref{zerocond}. Thus we can 
emphasize that the threshold bias voltage of the electron conduction through 
the moebius strip can be controlled very significantly by tuning the 
transverse hopping strength ($t_{\bot}$). From the results studied above 
we see that the behavior of the electron conduction is highly influenced 
by the peculiar topology of the moebius strip. 

The behavior of the electron conduction through the moebius strip is much 
more clearly visible by studying the current-voltage ($I$-$V$) 
characteristics. In the forthcoming part we shall discuss the characteristics 
of the current ($I$) and the noise power of its fluctuations ($S$) as a 
function of the applied bias voltage ($V$). Both the current ($I$) and the 
noise power of its fluctuations ($S$) are computed by the integration 
procedure of the transmission function $T$, as given in Eq.(\ref{equ5}) and 
Eq.(\ref{equ6}), respectively. The variation of the transmission function 
$T$ is similar to that of the conductance spectra, differ only in magnitude 
by a factor $2$ (since $g=2T$, from the Landauer conductance formula), as 
shown in Figs.~\ref{finitecond}, \ref{zerocond}, \ref{finitecondodd} 
and \ref{zerocondodd}.

In Fig.~\ref{finitecurr}, we draw the current and the noise power of its
fluctuations as a function of the bias voltage for the moebius strip  
in the limit of non-zero transverse hopping strength ($t_{\bot} \ne 0$). 
Here we set the total number of rungs $M=6$, same as in the study of 
Fig.~\ref{finitecond}. In this case ($t_{\bot} \ne 0$) the moebius strip is 
equivalent to a regular two-channel ring since the electrons can able to
move along the transverse direction. Figure~\ref{finitecurr}(a) 
and (b) correspond to the weak and the strong coupling cases, respectively,
where the solid curves correspond to the current and the dotted curves 
represent the noise power. For the weak coupling limit, the current shows
staircase-like behavior with sharp steps (solid curve of 
Fig.~\ref{finitecurr}(a)) associated with the sharp resonant peaks in the 
conductance spectrum (Fig.~\ref{finitecond}(a)), since the current is 
evaluated by the integration procedure of the transmission function $T$. 
On the other hand, as the coupling strength increases the staircase-like 
behavior disappears and the current varies continuously with the applied 
voltage $V$, as shown by the solid curve in Fig.~\ref{finitecurr}(b). This 
is due to the broadening of the resonant peaks in the conductance spectrum 
(Fig.~\ref{finitecond}(b)) for the limit of strong coupling. The another 
important point is that with the increase of the coupling strength, the 
current amplitude gets an order of magnitude enhancement which is clearly 
observed by comparing the results given by the solid curves in 
Figs.~\ref{finitecurr}(a) and (b). This feature can be understood by noting 
the areas under the curves those are plotted in Figs.~\ref{finitecond}(a) 
and (b), respectively. Both for the weak and strong coupling cases, we get 
finite value of the current momentarily as we switch on the bias voltage 
across this bridge. Now in the determination of the noise power of the 
current fluctuations (dotted curves in Fig.~\ref{finitecurr}), we see that 
both for the two coupling cases the shot noise achieves the sub-Poisson 
limit ($F<1$). Therefore, we can predict that the electrons are correlated 
with each other. Here the electrons are correlated only in the sense that 
one electron feels the existence of the other according to the Pauli 
exclusion principle (since we have neglected all other types of 
electron-electron interactions in our present formalism).

The characteristic properties of the current and the noise power of its 
fluctuations are much more interesting for the case where the electrons are
unable to hop along the transverse direction i.e., $t_{\bot}=0$. 
\begin{figure}[ht]
{\centering \resizebox*{8cm}{9.5cm}{\includegraphics{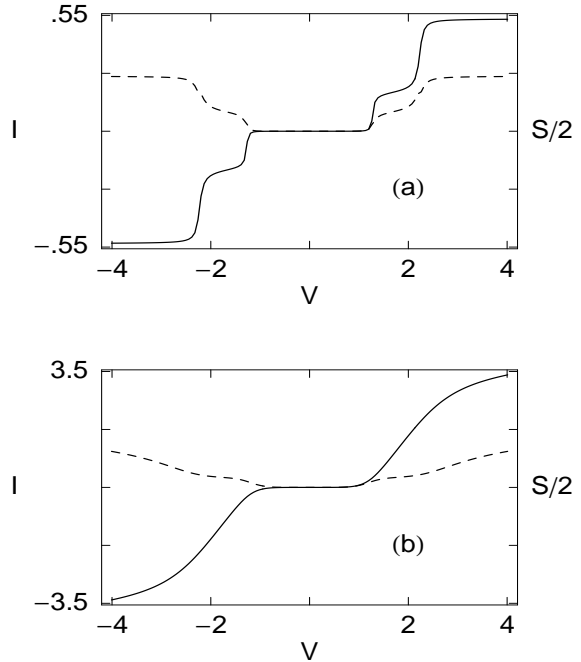}}\par}
\caption{Current $I$ (solid line) and the noise power of its fluctuations 
$S$ (dotted line) as a function of the voltage $V$ for the moebius strip 
with $t_{\bot}=0$. For this particular case the moebius strip is equivalent
to a single chain of length $2M$. Here we take $M=6$ (even) so that the 
length of the chain becomes $12$ in unit of the lattice spacing. (a) and 
(b) correspond to the results for the weak and the strong coupling limits, 
respectively.}
\label{zerocurr}
\end{figure}
For this particular case ($t_{\bot}=0$), as we have mentioned earlier, the 
moebius strip eventually reduces to a single chain with length $2M$. This
is due to the strange topological behavior of the moebius strip. 
Figure~\ref{zerocurr} shows the results for the moebius strip when 
$t_{\bot}=0$, where Figs.~\ref{zerocurr}(a) and (b) correspond to the results
for the weak and strong coupling cases, respectively. Here we take $M=6$,
so that the length of the chain becomes $12$, same as in the study of the
conductance behavior (plotted in Fig.~\ref{zerocond}). The solid and 
dotted curves of Fig.~\ref{zerocurr} represent the same meaning as in 
Fig.~\ref{finitecurr}. For the weak coupling limit, the current shows 
staircase-like behavior (solid curve of Fig.~\ref{zerocurr}(a)), while, 
it gets a continuous variation (solid curve of Fig.~\ref{zerocurr}(b)) in 
the limit of strong coupling as a function of the applied bias voltage, 
similar to the results as predicted for the strip with $t_{\bot} \ne 0$ 
(solid curves of Figs.~\ref{finitecurr}(a) and (b), respectively). The 
significant observation is that for this bridge (strip with $t_{\bot}=0$), 
both for the weak and strong coupling limits, the current appears after 
some critical value of the applied bias voltage (contrary to the results 
as given for the bridge with $t_{\bot} \ne 0$ i.e., for a regular 
two-channel ring (Fig.~\ref{finitecurr})). Thus the electron transport 
through the moebius strip can be tuned significantly by controlling 
the transverse hopping strength. A similar kind of behavior has also been
observed in the conductance spectra for the moebius strip with $t_{\bot} 
\ne 0$ and $t_{\bot}=0$ cases, respectively. Finally, in the study of the 
noise power of the current fluctuations
for this bridge we see that both for the two coupling cases the shot noise
goes from the Poisson limit ($F=1$) to the sub-Poisson limit ($F<1$) as long 
as we cross the first step in the current-voltage characteristics. Therefore, 
we can predict that the electrons are correlated after the tunneling process 
has completed in this bridge. This result is different from our previous 
studies i.e., the result for the strip with $t_{\bot} \ne 0$ where the shot
noise always lies in the sub-Poisson limit and there is no such possibility 
of transition from the Poisson limit to the sub-Poisson limit.

For the moebius strip with odd number of rungs we get quite similar features, 
as described above, in the current-voltage characteristics and also in the 
noise power of the current fluctuations. This is why here we do not describe 
these features further for the moebius strip with odd number of rungs.

\section{Concluding remarks}

To summarize, we have introduced a parametric approach based on the 
tight-binding model to investigate the electron transport properties
through a moebius strip attached to two metallic electrodes. The topological
effect of the moebius strip has an important signature in the electron
transport through such system. For the case where electrons are able to hop
along the transverse direction, the moebius strip becomes a regular 
two-channel ring. On the other hand, if the electrons are unable to hop 
along the transverse direction then the system reduces to a single chain 
with doubling its length than the previous one. Here we have described our 
results both for these two cases and obtained several interesting results.

For the weak-coupling limit, the conductance shows sharp resonant peaks 
(Figs.~\ref{finitecond}(a), \ref{zerocond}(a), \ref{finitecondodd}(a)
and ~\ref{zerocondodd}(a)) associated with the energy eigenvalues of the 
moebius strips. The widths of these resonant peaks become broadened 
substantially (Figs.~\ref{finitecond}(b), \ref{zerocond}(b), 
\ref{finitecondodd}(b) and \ref{zerocondodd}(b)) in the limit of strong 
coupling, where the contribution comes from the imaginary parts of 
the two self-energies, $\Sigma_S$ and $\Sigma_D$, due to coupling of the 
strip to the electrodes~\cite{datta}. Both for the moebius strips with
odd and even number of rungs, the threshold bias voltage of the electron
conduction can be tuned very nicely by controlling the transverse hopping 
strength $t_{\bot}$ which provides an important finding.

In the study of $I$-$V$ characteristics we have seen that the current shows 
staircase-like behavior with sharp steps (solid curves of 
Figs.~\ref{finitecurr}(a) and \ref{zerocurr}(a)) in the weak-coupling limit, 
while it varies continuously and achieves higher amplitude (solid curves
of Figs.~\ref{finitecurr}(b) and \ref{zerocurr}(b)) with the increase of 
the coupling strength. 

In the determination of the noise power of the current fluctuations, we have
noticed that for the strip with $t_{\bot} \ne 0$ the shot noise lies always
in the sub-Poisson limit ($F<1$) (dotted curves of Fig.~\ref{finitecurr}) 
i.e., the electrons are correlated with each other. On the other hand for 
the strip with $t_{\bot}=0$ the shot noise makes a transition from the 
Poisson limit ($F=0$) to the sub-Poisson limit ($F<1$) after the first step 
in the $I$-$V$ curve (dotted curves of Fig.~\ref{zerocurr}).

Throughout our discussions we have used several approximations by neglecting 
the effects of the electron-electron interaction, all the inelastic processes,
the Schottky effect, the static Stark effect, etc. More studies are expected 
to take into account all these approximations for our further investigations.

\end{document}